\documentclass[11pt,a4paper]{article}

\usepackage{jheppub}

\newcommand{\beq}{\begin{equation}}
\newcommand{\eeq}{\end{equation}}
\newcommand{\bea}{\begin{eqnarray}}
\newcommand{\eea}{\end{eqnarray}}
\newcommand{\ba}{\begin{array}}
\newcommand{\ea}{\end{array}}
\newcommand{\bc}{\begin{center}}
\newcommand{\ec}{\end{center}}
\newcommand{\bit}{\begin{itemize}}
\newcommand{\eit}{\end{itemize}}
\newcommand{\eq}[1]{Eq.~(\ref{#1})}

\def\htm{h \to \tau^\pm  \mu^\mp }
\def\tm{ \tau^\pm  \mu^\mp }
\def\mt{ \tau^\pm  \mu^\mp }
\def\gg{\gamma \gamma }
\def\cbma{\cos (\beta - \alpha)}
\def\sbma{\sin (\beta - \alpha)}

\def\tmg{\tau \to \mu \gamma}
\def \gsim{\mathrel{\vcenter
     {\hbox{$>$}\nointerlineskip\hbox{$\sim$}}}}

\makeatletter
\def\@fpheader{\relax}
\makeatother

\title{Two Higgs doublets to explain the excesses $pp\rightarrow \gamma\gamma(750\ {\rm GeV})$ and $\htm$}

\author{Nicolas Bizot$^{1}$,}
\author{Sacha Davidson$^{2}$,}
\author{Michele Frigerio$^{1}$,}
\author{Jean-Lo\"ic Kneur$^{1}$}

\affiliation{$^{1}$Laboratoire Charles Coulomb (L2C), UMR 5221 CNRS-Universit\'e de Montpellier, \\
Place Eug\'ene Bataillon, F-34095 Montpellier, France}

\affiliation{ 
$^{2}$IPNL, CNRS/IN2P3,  4 rue E. Fermi, 69622 Villeurbanne cedex, France; 
Universit\'e Lyon 1, Villeurbanne;
 Universit\'e de Lyon, F-69622, Lyon, France}

\abstract{ The two Higgs doublet model emerges as a minimal scenario in which to address, at the same time, the $\gamma\gamma$ excess at 750 GeV and the lepton flavour violating decay into $\tm$ of the 125 GeV Higgs boson.
The price to pay is additional matter 
to enhance the  $\gamma\gamma$ rate, and a peculiar pattern for the lepton Yukawa couplings.
 We add TeV scale vector-like fermions and find parameter space
consistent with  both excesses, as well as with Higgs and electroweak precision observables.}

\begin{document}
\maketitle
\flushbottom

\section{Introduction}

The recently presented indications for a diphoton excess at ATLAS and CMS at an invariant mass of 750 GeV \cite{run2} have caused much excitement in the high-energy phenomenology community \cite{FGS,notjust2hdm,doublet,VLrun,2bumps,axion,scalar,nomura,composite,vector,spin2,dm,modindep,other,SUSY,flavour}.
 At the same time, some hints of anomalies persist in the LHC run-I data. Notably there is a $2.4\sigma$ excess at CMS in the $\htm$ decay
of  the 125 GeV Standard Model-like Higgs boson $h$  \cite{CMShtm}, corresponding to a best-fit branching ratio BR$(\htm)=0.84^{+0.39}_{-0.37}\,\%$. This is compatible with the ATLAS analysis which finds BR$(\htm)=0.77\pm 0.62\,\%$ \cite{ATLAShtm}.

One of the simplest renormalisable models allowing for a $\htm$ branching ratio of the order of a percent is the two-Higgs-doublet model (2HDM) \cite{2HDMrev} with lepton flavour violating (LFV) Yukawa couplings. This model has been studied before \cite{bjw,Grenier,oldLFV}, and, with renewed interest, after the CMS excess was announced \cite{newLFV}. 
The aim of the present paper is to study whether a simple 2HDM 
could explain both the diphoton excess and the LFV Higgs decay. 
The LHC excess has been studied  in the
2HDM in \cite{notjust2hdm,doublet}, and several authors
have  combined it   with other
observables, such as  the dark matter abundance \cite{dm},
or $B$-physics anomalies \cite{flavour}.
In the 2HDM, candidates for the 750 GeV resonance are the heavier  scalar 
Higgs $h_2$ and the pseudoscalar $A$. They can reproduce the observed cross-section times branching ratio 
into photons if they couple to heavy vector-like charged fermions, as has been discussed by several 
authors, e.g. \cite{FGS,notjust2hdm,VLrun}. 
The data \cite{run2} suggest a broad resonance, which
could be due to the exchange of nearly degenerate $h_2$ and $A$ \cite{2bumps}.

We consider a CP-conserving 2HDM  of type I in the decoupling limit \cite{GunionHaber},
where the second doublet  has a mass $\sim 750$ GeV.  
 We work in
the ``Higgs basis'', 
 where  $H_1 = [0,(v+h_1)/\sqrt{2}]$ denotes the doublet which gets
a vacuum expectation value (vev) $v\simeq 246$ GeV, and  which has 
Standard Model Yukawa couplings. The second doublet $H_2= [H^+,(h_2 + i A)/\sqrt{2}]$
does not couple to Standard Model fermions, except for a LFV Yukawa to $\tm$.
The physical Higgs bosons are the CP-even $h$ and
$H$, the pseudoscalar $A$ and the charged Higgses
$H^\pm$.
In the decoupling limit,
the  light   $h$
 is almost aligned on the vev, making it 
the Standard-Model-like Higgs
of 125 GeV.
In section 2 we show how to enhance the $H$ and $A$ couplings
to gluons and photons, by introducing new vector-like charged fermions, while respecting the bounds
from electroweak precision tests and $h$ signal strengths.
We neglect the charged Higgs $H^+$ because it contributes little to $H,A \to \gg$.
A small mixing with $h_2$ allows the LFV decay $h\to\mt$, as
discussed in section \ref{sec:LFV}.
In section 4, we demonstrate that one can accommodate the $750$ GeV excess from the decays of $H$ and $A$,
in agreement with the LFV excess.

\section{Two Higgs doublets coupling to extra matter}\label{HF}

In this section we
 neglect the misalignment between the CP-even
mass basis, and the ``Higgs'' basis, and focus on the
Higgs couplings to new fermions. That is, we
 consider the limit where the Standard Model Higgs boson $h$ is identified with $h_1$, and the second Higgs doublet $H_2$ does not couple to the Standard Model, except for its gauge interactions.
Therefore, $H=h_2$ and $A$ cannot decay to Standard Model particles at tree-level.
We  include the
misalignment  in the following section, in order to 
obtain $\htm$.

In order for $H$ and/or $A$ to play the role of the $750$ GeV resonance, we need to introduce a large effective coupling to $\gamma\gamma$, as well as to $gg$, in the hypothesis that the resonance is produced via gluon fusion.
If the production is dominated by quarks, that have a smaller parton density function,  one needs an even larger coupling to $\gamma\gamma$. 
 We will discuss quantitatively these two possibilities in section 4.

To provide an explicit realization for such effective couplings, we introduce two vector-like fermions, that transform under $SU(3)_c\times SU(2)_w\times U(1)_Y$ as $D\sim (R_c,2,Q+1/2)$ and $S\sim(R_c,1,Q)$, with interactions
\beq
-{\cal L} = M_D \overline{D_L} D_R + M_S \overline{S_L} S_R + \lambda^D_i \overline{D_L} H_i S_R + \lambda^S_i \overline{S_L} H_i^\dagger D_R +h.c. ~.
\label{lag}\eeq 
The state of electric charge $Q+1$ has mass $M_D$ and no Yukawa couplings. The two states of charge $Q$ couple to the Higgs doublets,
and their mass matrix is non-diagonal because of the vev of $H_1$. We will denote the mass eigenvalues by $M_1\le M_2$.
Note that, in order to induce the couplings $H\gamma\gamma$ and $A\gamma\gamma$, one needs both $\lambda^{D,S}_2\ne 0$ and $\lambda^{D,S}_1\ne 0$,
to generate the effective operator $H_2^\dagger H_1 F_{\mu\nu}F^{\mu\nu}$ via a fermion loop (and analogously for gluons).

The couplings $\lambda^{D,S}_1$ are constrained as they contribute to the $h$-decays into $\gamma\gamma$ and $gg$, as well as by the precision electroweak parameters $S$ and $T$.
Indeed, vector-like charged fermions were employed in the past to explain the transient excess in the $h\to\gg$ channel, see e.g. \cite{oldHiggsPhotonPhoton}.
A detailed analysis of the allowed parameter space is provided in Ref.~\cite{Bizot:2015zaa}.  Here we describe two illustrative cases:
\bit 
\item[(1)] Degenerate fermion masses, $M_1=M_2$. This is the case for
$M_D=M_S$ and  $\lambda^S_1=-\lambda^D_1$. Choosing $M_{1,2}=1$ TeV, $R_c=3$ and $|Q|\le 2$, one finds an upper bound 
$(\lambda^S_1 v) /(\sqrt{2} M_1) \lesssim 0.25$. 
This bound is determined essentially by the $T$ parameter, that is proportional to $N_c\equiv\dim(R_c)$ and independent from $Q$. When $|Q|>2$ a stronger bound comes from the Higgs signal strengths.
For $R_c=8$ and $|Q|\le 3$, one needs $(\lambda^S_1 v) /(\sqrt{2} M_1) \lesssim 0.12$. In this case the bound comes from the $hgg$ coupling.
\item[(2)] One vanishing Yukawa coupling, e.g. $\lambda^D_1=0$. 
This pattern strongly suppresses the correction to the couplings $h\gamma\gamma$ and $hgg$,
because,  in the limit of heavy fermions, they are proportional to $\lambda^D_1\lambda^S_1$. 
However, an upper bound on $\lambda^S_1$ still exists, coming from the $T$ parameter, $(\lambda^S_1 v) /(\sqrt{2} M_1)  \lesssim 0.35~(0.25)$ for $R_c=3~(8)$ and $M_1=1$ TeV. 
Note that $T$ does not depend on the hypercharge, therefore it turns out that one can take it very large, say $Q\sim 10$, without violating the constraints.
\eit

Let us now turn to the heavy Higgs doublet $H_2$. Its couplings to the fermion mass eigenstates are easily derived \cite{Bizot:2015zaa} 
in terms of the parameters in the Lagrangian of \eq{lag}. Then, one can compute the decay width into two photons for the scalar $H$ and the pseudoscalar $A$.
The result is particularly compact in the limit $M_{H} \ll 2 M_{1,2}$, since in this case the loop form factor $A_{1/2}[M^2_{H}/(4M_i^2)]$ is the same for both fermions
in very good approximation, $A_{1/2}(0)=4/3$. Similarly, for $A$ we use the loop form factor $\tilde A_{1/2}(0)=2$. Then, one obtains
\beq
\frac{\Gamma(H\rightarrow \gamma\gamma)}{M_{H}}= \frac{\alpha^2}{256\pi^3}
\left|\frac{2v M_{H}}{3M_1M_2} N_c Q^2  (\lambda_2^D\lambda_1^S+\lambda_2^S\lambda_1^D)\right|^2 ~,
\label{2.2}\eeq
\beq
\frac{\Gamma(A\rightarrow \gamma\gamma)}{M_{A}}= \frac{\alpha^2}{256\pi^3}
\left|\frac{v M_{A}}{M_1M_2} N_c Q^2  (\lambda_2^D\lambda_1^S - \lambda_2^S\lambda_1^D)\right|^2 ~,
\label{2.3}\eeq
In the same approximation, the widths into two gluons read
\beq
\frac{\Gamma(H\rightarrow gg)}{M_{H}}= \frac{\alpha_s^2}{32\pi^3}
\left|\frac{2v M_{H}}{3M_1M_2} C(R_c)  (\lambda_2^D\lambda_1^S+\lambda_2^S\lambda_1^D)\right|^2 ~,
\label{2.4}
\eeq
\beq
\frac{\Gamma(A\rightarrow gg)}{M_{A}}= \frac{\alpha_s^2}{32\pi^3}
\left|\frac{v M_{A}}{M_1M_2} C(R_c)  (\lambda_2^D\lambda_1^S  -\lambda_2^S\lambda_1^D)\right|^2 ~,
\label{2.5}
\eeq
where $C(R_c)$ is the index of the color representation. Note that 
the ratio of $H$-rates over $A$-rates is given by a factor $(2|\lambda_2^D\lambda_1^S+\lambda_2^S\lambda_1^D|)^2/
(3|\lambda_2^D\lambda_1^S-\lambda_2^S\lambda_1^D|)^2$.

For definiteness, consider the case (2) described above, $\lambda^D_1=0$, and take
$M_{H}\simeq M_A\simeq 750$ GeV. 
Then, one obtains
\beq
\frac{\Gamma_{A\gamma\gamma}}{M_{A}}\simeq \frac 94 \frac{\Gamma_{H\gamma\gamma}}{M_{H}} \simeq 1.4\cdot 10^{-6} 
\left(\frac{1~{\rm TeV}}{M_2}\right)^2 \left(\frac{N_c}{3}\right)^2 \left(\frac Q2\right)^4 \left(\frac{\lambda_2^D}{3}\right)^2 ~,
\label{ppR}\eeq
\beq
\frac{\Gamma_{Agg}}{M_{A}}\simeq \frac 94 \frac{\Gamma_{Hgg}}{M_{H}} \simeq 4.4 \cdot 10^{-6} 
\left(\frac{1~{\rm TeV}}{M_2}\right)^2 \left(\frac{C(R_c)}{1/2}\right)^2  \left(\frac{\lambda_2^D}{3}\right)^2 ~,
\label{ggR}\eeq
where we chose $(\lambda^S_1 v) /(\sqrt{2} M_1) \simeq 0.35$, that is the largest value allowed by the $T$ parameter for $R_c=3$.
In the case of a colour octet, $N_c=8$ and  $C(R_c)=3$, there is a slightly stronger upper bound,
$(\lambda^S_1 v) /(\sqrt{2} M_1) \simeq 0.25$: therefore,
one gains a factor $\sim3$ in $\gamma\gamma$ and a factor $\sim20$ in $gg$.

Note that one can reproduce the same rates with smaller Yukawa couplings:  
taking $N$ pairs of vector-like fermions, all with equal charges and coupling $\lambda_2^{D}$, the rates scale as $(N \lambda_2^{D})^2$.
From a theoretical point of view, it may be more justified to introduce several vector-like fermions, but with charges related to the Standard Model ones, such as one or more
vector-like families, composed of $t,b$ and $\tau$ partners. Adding over their contributions one could obtain a qualitative similar effect.

One should also remark that the heavy fermion loops also induce decays of $H$ and $A$ to $Z\gamma$, $ZZ$ and $WW$, with width of the same order as (or slightly smaller than) for $\gamma\gamma$.
However, the upper bounds from the 8 TeV LHC are weaker than the one on  $\gamma\gamma$, as discussed e.g. in Ref.~\cite{nomura}. Therefore, they are presently unconstraining. 
At run 2, the better perspective appears to be the observation of the $Z\gamma$ channel.

\section{The $\mt$ decay of the 125 GeV Higgs boson}\label{sec:LFV}

Flavour-changing Higgs couplings are generic
in the 2HDM, but their effects are not seen
in  low energy precision experiments  searching for  lepton or quark 
 flavour change. So  a discrete symmetry,
which forbids flavour-changing Yukawa couplings,
is usually imposed on the 2HDM.  
To allow for LFV $h$ decays, without generating
undesirable flavour-changing processes, we 
suppose
that our 2HDM almost has a discrete symmetry:
all the Standard Model  fermions have
the usual Yukawa couplings to $H_1$ (``type I'' model),
and the only two couplings of $H_2$
to Standard Model fermions are the $\mu\tau$ LFV ones,
\beq
{\cal L}=- \rho_{\tau \mu} \overline{L_\tau } H_2  \mu_R
- \rho_{\mu \tau} \overline{L_\mu } H_2  \tau_R
+{\rm h.c.}\,
\eeq
(see   \cite{Howie,Grenier} for a more
formal analysis). By definition, these
LFV couplings are attributed to the doublet
$H_2$ with zero vev, because 
diagonalising fermion mass matrices diagonalises
the Yukawa couplings of $H_1$,  which carries the vev. 
In section 4 we will also consider a scenario where $H_2$ is produced from
an additional Yukawa coupling to $b$ quarks, that can be added
without phenomenological problems.

The  CP-even mass eigenstates $h$ and $H$ 
are misaligned with respect to the vev by
an angle that is commonly
 parametrized as  $\beta -\alpha$:
\beq
\ba{rcl}
h &=& \sbma h_1 + \cbma h_2 ~, \\
H &=& \cbma h_1 - \sbma h_2 ~. \\
\ea
\label{mis}
\eeq
In the decoupling
limit \cite{GunionHaber,Howie}, $\sbma \simeq 1$  and 
\beq
\cbma = -\frac{\Lambda_6 v^2}{M_H^2} ~,
\label{cbma}
\eeq
where the Higgs potential  contains a term
$\Lambda_6 H_2^\dagger H_1 H_1^\dagger H_1 + h.c.$,
in the basis where $H_1$ has no vev.
The coupling of
$h$  to $\tm$  is therefore  proportional to $\cbma \rho$, and one obtains
\beq
BR (\htm) \simeq \frac{m_h}{16 \pi \Gamma_h} 
\cos^2(\beta-\alpha){\Big (}|\rho_{\tau \mu}|^2 + |\rho_{\mu \tau}|^2  {\Big )} ~.
\eeq
The CMS best-fit is $BR(\htm) = 0.0084$ \cite{CMShtm}, which gives
\beq
\cos(\beta-\alpha)(|\rho_{\tau \mu}|^2 + |\rho_{\mu \tau}|^2)^{1/2}\simeq 0.0037 ~,
\label{LFVnumber}\eeq
where the width was taken  at its Standard Model
value, $\Gamma_h \simeq 4.1$ MeV. 

In the 2HDM, the CMS excess in $\htm$ is consistent with the
current upper bound  
$BR(\tau \to \mu \gamma) \leq 2.6\times 10^{-7} 
BR(\tau \to \mu \nu \bar{\nu})$ \cite{BelleBabar}.
However,  the extra fermions which enhance
$H, A \to \gg$ as in Eqs.~(\ref{2.2})-(\ref{2.3}), will also
enhance the rate for $\tmg$ \cite{bjw}:  if  a
neutral Higgs is exchanged between its
$\gg$ and $\bar{\tau}\mu$ vertices, and
one of the photons connects to the lepton line,
a diagram for $\tmg$ is obtained.
Such diagrams with a top loop were
calculated in the 2HDM  in \cite{CHK}. From their
results,  the  combined 
contribution  of  $H$ and $A$ can be estimated, for 
$M_1 \simeq M_2$ and $\lambda_1^D = 0$,  as
\beq
\frac{ m_\tau}{v^2} A_L \simeq \frac{e\alpha}{128\pi^3}\frac{v}{\sqrt{2} M^2_1}
N_cQ^2 \lambda_2^D \lambda_1^S \rho^*_{\tau \mu} ~,
\label{BZestimate}
\eeq
where the
experimental bound is  $ 384\pi^2 (A_L^2 + A_R^2) \leq 2.6 \times 10^{-7}$.
 With the  definition of Yukawa couplings given in
Eq. (\ref{lag}), it turns out that  choosing a large
$\lambda_2^D$ ($\lambda_2^S$) leads to a destructive (constructive) interference among the $H$ and $A$ amplitudes.
This was taken into account in \eq{BZestimate}, 
where  the difference in  loop integral functions 
was chosen $\simeq 1/2$, 
as given  in \cite{CHK} for $M_1^2/M_H^2 \simeq 2$. 
A similar estimate can be made for $A_R$.
We neglect the $h$ contribution to
$\tmg$, because its coupling to $\gg$
is not enhanced, see scenario (2) in
section \ref{HF}. 
So the Babar-Belle bound on $\tmg$  could be
satisfied for 
\beq
\frac{N_c}{3} \left(\frac Q2\right)^2 
\frac{\lambda_2^D }{3}  \lambda_1^S \rho_{ \tau \mu  }
\lesssim  0.07 ~,
\label{NCbig}
\eeq
which sets a lower bound on $\cbma$ when combined
with  Eq. (\ref{LFVnumber}): 
\beq
 \cos^2(\beta-\alpha) \gsim  0.003 \left(\frac{N_c}{3}\right)^2 \left(\frac Q2\right)^4 
\left(\frac{\lambda_2^D }{3}\right)^2  \left(\lambda_1^S\right)^2 ~~.
\label{LFVnumber2}
\eeq
If the masses  and couplings were purposefully tuned,
it might be possible to suppress
the $\tmg$ amplitude even further, so
we will  consider
 Eq. (\ref{LFVnumber2}) to be a preference 
but not an exclusion.

\section{Reproducing the 750 GeV excess} 

Let us discuss the decay widths of $H$ and $A$ as a function of the Higgs mixing $\cos(\beta-\alpha)$ and of the LFV couplings $\rho_{\mu\tau,\tau\mu}$.

The mixing does not affect the couplings of the pseudoscalar $A$, for which the discussion of  section \ref{HF}  applies.
On the other hand, the misalignment
parametrised in \eq{mis} implies
that the Yukawa couplings to $h$ and $H$ become 
\beq
\lambda^{D,S}_h = \sbma \lambda^{D,S}_1 +\cbma \lambda^{D,S}_2~,~~~~~
\lambda^{D,S}_H = \cbma \lambda^{D,S}_1 -\sbma \lambda^{D,S}_2~.
\eeq
The $H$ decay widths into photons and gluons are obtained
by replacing  $\lambda^{D,S}_2$ with  $\lambda^{D,S}_H$ in Eqs.~(\ref{2.2}) and (\ref{2.4}).
Similarly, for the corrections to $h\to \gg$ and $h\to gg$ due to the heavy fermions, one has to replace $\lambda^{D,S}_1$ with  $\lambda^{D,S}_h$.
In addition, all the $h$ couplings to the Standard Model particles $n$'s are modified, 
$g_{hn\bar n} = \sin(\beta-\alpha) g^{SM}_{hn\bar n}$. 
Since several Higgs signal strengths have been tested at LHC-8 TeV with 10\% precision, the Higgs mixing is bounded from above
\beq
\cos^2(\beta-\alpha) \lesssim 0.1 ~.
\label{cbmaU}\eeq 
This is  consistent with Eq.~(\ref{LFVnumber2}).
As discussed in section \ref{HF}, the corrections to $h\to \gg$ and $h\to gg$ 
may lead to a slightly stronger upper bound on $\cbma$, if the couplings $\lambda^{D,S}_2$ are very large. 
However, such bound drops for $\lambda_1^S\cdot \lambda_1^D\to 0$, see case (2) in section \ref{HF}.
Finally, the contributions to $S$ and $T$ from scalar loops are small in the 2HDM close to the decoupling limit \cite{Grenier,Baak:2011ze},
as we explicitly checked for our choice of the parameters.

The mixing has an important effect on the total width of $H$, since the latter can decay to Standard Model particles $n$'s, 
with coupling $g_{Hn\bar n} = \cos(\beta-\alpha) g^{SM}_{hn\bar n}$. 
The dominant contributions read, at the tree level,
\beq
\dfrac{\Gamma(H\rightarrow t\bar t,W^+W^-,ZZ)}{M_H} 
\simeq 
\dfrac{ \cos^2(\beta-\alpha)}{8\pi\,v^2} \left[ 3 m_t^2 
+ \dfrac{M^2_H}{2}  + \dfrac{M_H^2}{4} \right] \\
\simeq
0.33 \cos^2(\beta-\alpha)~,\\
\label{SM}
\eeq
where, for the latter numerical estimate, we used the accurate values of the widths for $M_H\simeq 750$ GeV, as given in Ref.~\cite{Dittmaier:2011ti}.
Here we neglected the channel $H\to hh$, because the corresponding trilinear scalar coupling may be suppressed, by conveniently choosing the scalar potential parameters.
Recall that the cross-section for $pp\rightarrow H\rightarrow \gamma\gamma$ is proportional to $\Gamma(H\to gg)/\Gamma_H^{tot}$,
where the numerator corresponds to the assumed dominant $H$ production mode, and the denominator is the total width of $H$. 
Therefore, the contribution of $H$ to the excess degrades as soon as $\Gamma(H\rightarrow gg)/M_H \lesssim 0.33 \cos^2(\beta-\alpha)$.

The LFV couplings $\rho_{\mu\tau,\tau\mu}$ also open an additional decay channel for both $H$ and $A$, with a width
\beq
\frac{\Gamma(H\rightarrow \tm)}{M_H}\simeq \frac{\Gamma(A\rightarrow \tm)}{M_A}\simeq
\frac{1}{16 \pi} 
{\Big (}|\rho_{\tau \mu}|^2 + |\rho_{\mu \tau}|^2  {\Big )} \simeq \frac{3 \cdot 10^{-7}}{\cos^2(\beta-\alpha)}~,
\label{mtR}\eeq
where the last equality comes from \eq{LFVnumber}.

One should also mention that the presently preferred width of the excess, $\Gamma \sim 45$ GeV,
could be mimicked by two narrow resonances close in mass.
Indeed, the mass split between $H$ and $A$ is given, in the decoupling limit, by  $M^2_{H}-M^2_A \simeq \Lambda_5 v^2 $,
where the term $\frac{1}{2}\Lambda_5 (H_1^\dagger H_2)^2 + h.c.$ appears in the Higgs
potential. This is naturally of the correct order of magnitude for $\Lambda_5\simeq 1$.
Note, however, that the $H$-mediated cross-section tends to be suppressed relatively to the $A$-mediated one by two factors:
the additional Higgs width in \eq{SM}, and the factor $4/9$ from the loop form factors, see Eqs.~(\ref{ppR})-(\ref{ggR}).

Let us put all the constraints together to identify the possible windows of parameters that allow to reproduce the $750$ GeV excess
in agreement with the preferred $h\to\tm$ rate. 
The resonant LHC total cross-section, in the 
crude zero-width approximation, reads 
\beq
\sigma (pp \to H(A) \to \gamma\gamma) = \sum_i P_i\:
\frac{\Gamma(H(A)\to i) \Gamma(H(A)\to \gg)}{s\:\Gamma_{tot} M_{H(A)}}~,
\label{sigNWA}
\eeq
where 
$s=(13 {\rm~TeV})^2$, $M_{H(A)}\simeq 750$ GeV, and 
the $P_i$ coefficients are the integrals for convoluting over parton densities, that define the parton luminosities 
for each species $i$: 
\bea
&& P_{gg} \equiv \frac{\pi^2}{8}  \int ^1_{\frac{M^2}{s}} \frac{d\,x}{x} g(x) g(\frac{M^2}{x\,s}) , \nonumber \\
&& P_{\bar q q} \equiv \frac{4\pi^2}{9}  \int ^1_{\frac{M^2}{s}} \frac{d\,x}{x} 
\left[ q(x) \,\bar q(\frac{M^2}{x\,s}) +\bar q(x)\, q(\frac{M^2}{x\,s}) \right] . 
\label{pdf}
\eea
Consistency with the absence of resonances at 8 TeV favours $i$ to be
either gluons or $b$s, for which the luminosity is $P_{\bar b b} \simeq 14$
and $P_{gg}\simeq 2000$ (we used for  \eq{pdf} the latest pdfs from Ref.~\cite{CTeQlast}).

\begin{figure}[bt]
\begin{center}
\includegraphics[width=15cm]{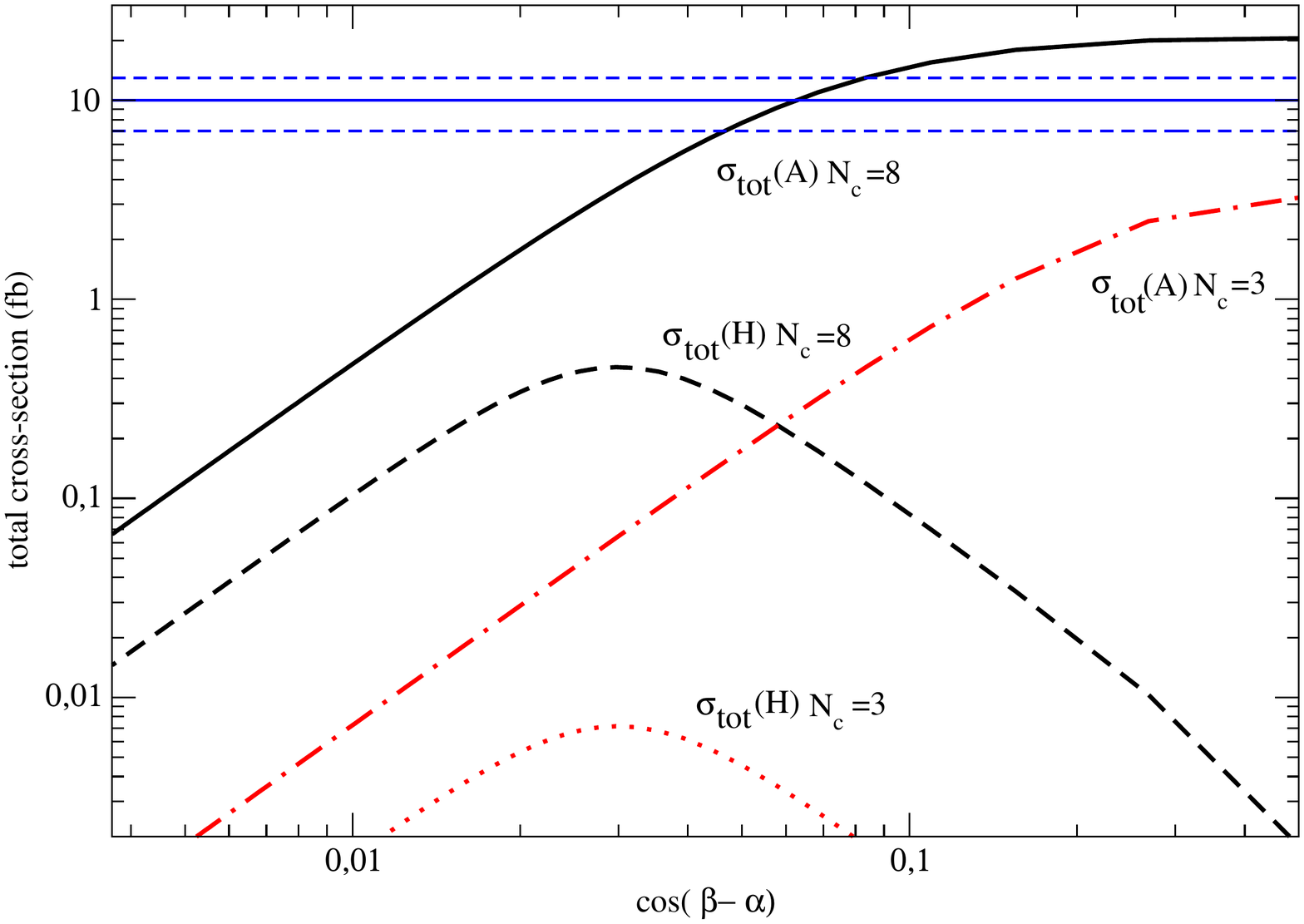}
\caption{The total cross-section $\sigma(pp\rightarrow H(A) \rightarrow \gg)$ in fb, assuming the gluon fusion production channel, 
as a function of $\cos(\beta-\alpha)$, for a pair of vector-like fermions in the color representation $R_c=3$ or $R_c=8$, as indicated. 
We fixed their charge, $Q=2$, and their Yukawa couplings to $H$ and $A$, $\lambda^D_2=3$ and $\lambda^S_2=0$. 
The horizontal band is the preferred cross-section at $1\sigma$
for the ATLAS excess \cite{run2}.
\label{fig1}}
\end{center}
\end{figure}

We focus first on gluon-gluon fusion as the dominant production mechanism. 
This channel enjoys the largest parton density functions, so it is sufficient to have $\Gamma(H,A\to\gg)/M_{H,A}\simeq 10^{-6}$  \cite{FGS}, as long
as $\Gamma^{H,A}_{tot}\simeq \Gamma(H,A\to gg)$. However, the latter is loop-suppressed as shown in \eq{ggR}.
The total cross-section for some choices of the parameters is shown in Figure \ref{fig1} as a 
function of $\cbma$. Note that
for completeness and cross-check, we have also compared with the more elaborated invariant mass distribution
$d\sigma/dM(g g\to H(A)\to \gg)$ 
where $M\equiv \sqrt{\hat s}$ is the $\gg$ invariant mass, that we have calculated
taking into account the exact width dependence, 
and integrating this expression over an appropriate large range for $M$ around the resonance. The numerical 
differences with the narrow-width approximation expression in \eq{sigNWA} is at most 2-3 \%
for all the relevant parameter choices discussed below, as could be intuitively expected since 
the total width of either $A$ or $H$ remains in all cases
sufficiently moderate with respect to the resonance mass, such that the narrow width approximation is
justified a posteriori. 
We can envisage two scenarios:
\bit
\item 
 [(A)] For both $H$ and $A$ to contribute significantly to the excess, one has to compete with both the tree-level widths in \eq{SM} and \eq{mtR}.
So  the optimal value for the Higgs mixing is 
$\cos^2(\beta-\alpha)\simeq 10^{-3}$. Then, to reach a cross-section of a few fbs
one needs $\Gamma(H,A\to gg)\Gamma(H,A\to \gg)/M_{H,A}^2\gtrsim 5\cdot 10^{-10}$.
To reach this value for both $H$ and $A$ requires some stretch in the parameters, e.g. in Eqs.~(\ref{2.2})-(\ref{2.5}) 
one should take $R_c=8$, $Q=3$, $\lambda_2^S=-1$, $\lambda_2^D=5$, $M_1=M_2$ and $\lambda_1^S=-\lambda_1^D$ with the corresponding constraint 
$(\lambda^S_1 v) /(\sqrt{2} M_1) \lesssim 0.12$.
 In addition, the amplitude for $\tmg$  in
 this scenario exceeds the indicative bound  of \eq{NCbig} 
by about an order of magnitude.
\item [(B)] If one renounces to the $H$ contribution to the excess and focuses on $A$, only the width in \eq{mtR} competes with gluon fusion.
One can take the Higgs mixing as large as allowed by 
Standard Model constraints, $\cos^2(\beta-\alpha)\simeq 0.1$ (see \eq{cbmaU}).
Then, one can reach a cross-section of a few fbs as long as 
 $\Gamma(A\to gg)\Gamma(A\to \gg)/M_A^2\gtrsim  3\cdot 10^{-12}$, as realized with the reference values in Eqs.~(\ref{ppR})-(\ref{ggR}).
The bound (\ref{LFVnumber2})  from $\tmg$   
is satisfied for these parameters.
\eit

\begin{figure}[bt]
\begin{center}
\includegraphics[width=15cm]{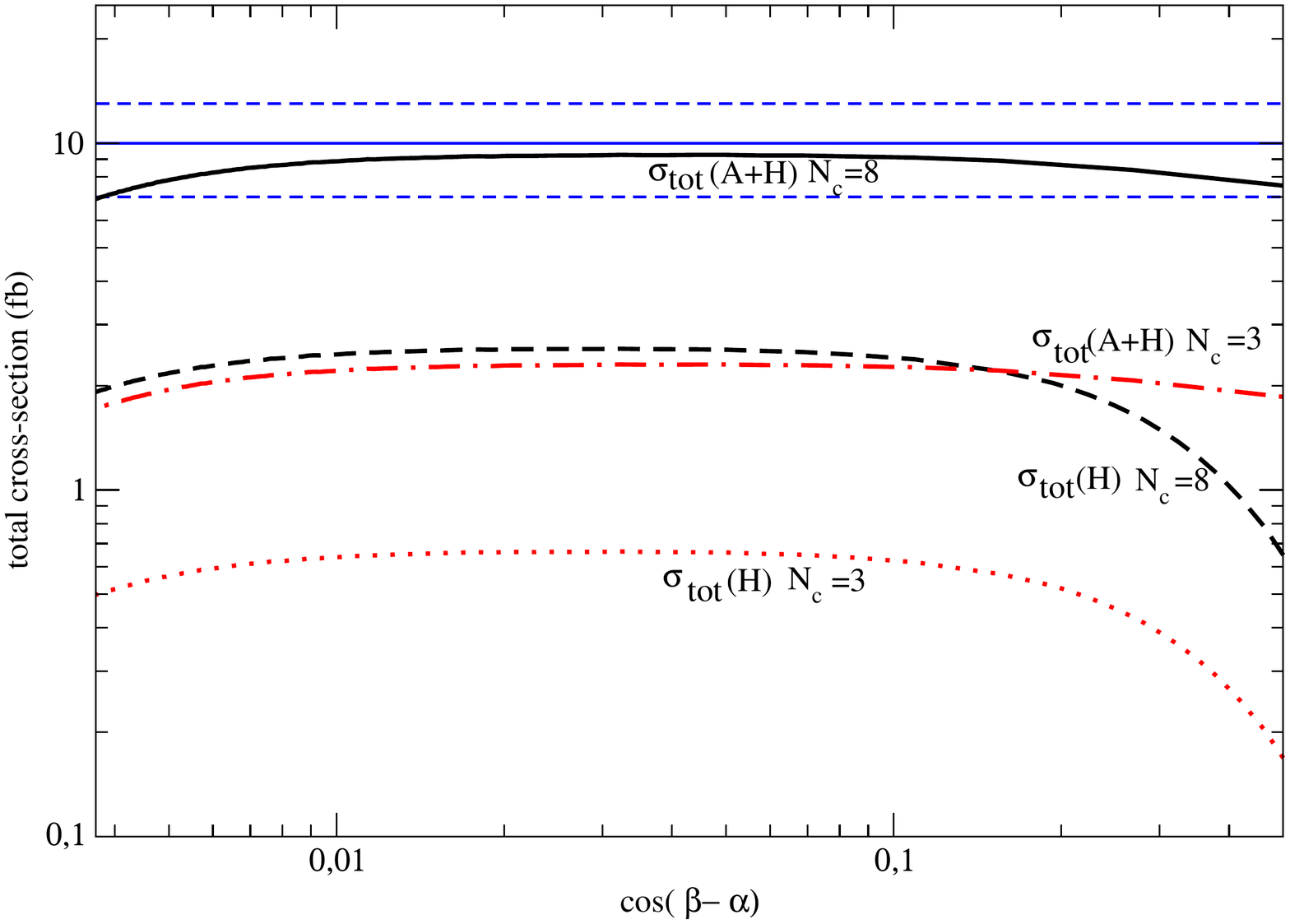}
\caption{The same as in Fig. 1, but adding 
a $b\bar b$ production channel with $\rho_b =1 $, see \eq{bbwidth},  
and increasing the vector-like fermion charge, $Q=5$.
Here we displayed the cross-section for $H$ only, as well as the cross-section for $H$ plus $A$.
\label{fig2}}
\end{center}
\end{figure}

Let us compare with the alternative possibility that the production of $H$ and $A$ is not dominated by gluon fusion, rather by $b\bar b\to H,A$. 
The parton density functions give a suppression of order 100 with respect to gluons, so that the excess demands for $\Gamma(H,A\to\gg)/M_{H,A}\gtrsim {2\cdot 10^{-4}}$ \cite{FGS}.
The advantage is that a Yukawa coupling $(\rho_b/\sqrt{2}) \overline b (h_2+i\gamma_5A) b$ can easily overcome the other tree-level widths in \eq{SM} and \eq{mtR},
\beq
\dfrac{\Gamma(H\rightarrow b\bar b)}{\sin^2(\beta-\alpha)M_H} \simeq
\dfrac{\Gamma(A\rightarrow b\bar b)}{M_A} \simeq 
\dfrac{ 3 \rho_b^2}{16\pi}  \simeq 0.06 \rho_b^2 ~.\\
\label{bbwidth}\eeq
Indeed, one can reproduce the preferred value $\Gamma\simeq 45$ GeV for $\rho_b\simeq1 $.
Moreover, there is no constraint from dijet searches at 8 TeV, as the $b$-quark parton density function is very small.
Therefore, one identifies the following scenario:
\bit
\item [(C)] When $\Gamma^{H,A}_{tot}\simeq \Gamma(H,A\to b\bar b)$, both $H$ and $A$ contribute to the excess, as long as $\Gamma(H,A\to\gg)\simeq { 2\cdot 10^{-4}}$.
Confronting with \eq{ppR}, one needs a pair of vector-like fermions with $R_c=3$ and $Q=7$, or $R_c=8$ and $Q=5$. Note that is difficult
to avoid such large exotic charges by augmenting the number of multiplets in the loop, as the signal scales with $Q^4$. 
As discussed in section \ref{HF}, such 
 large $Q$
can be compatible with Higgs decays and the  $S$ and $T$ parameters,
however
the bound  of Eq. (\ref{LFVnumber2}) from $\tmg$ is exceeded
by a factor of few.
\eit 
The total cross-sections, combining 
both the gluon fusion and $b\bar b$ production channels, are shown in Figure \ref{fig2} 
as a function of $\cbma$, for $Q=5$ 
and other parameters as in Figure \ref{fig1}. 
Here the cross-sections are calculated with the exact width dependence and 
integrating $d\sigma/dM(g g, b\bar b \to H(A)\to \gg)$. In fact due to the dominant contribution
of the $b\bar b$ decay to the total width $\Gamma_{tot}$ in this case, the  $b\bar b$ production channel 
largely dominates (for instance the gluon fusion process contributes to the total
cross-section by about $\sim 10\%$
only for $R_c=8$, and much less for $R_c=3$).  
Note that in this case the discrepancy with the cross-sections in the  
narrow width approximation of \eq{sigNWA} amounts to 
7-8 \%, for the parameter choices discussed above, that is roughly of order $\Gamma_{tot}^{H,A}/M_{H,A}$. 

\section{Final comments}

We entertained the possibilities that both the $\gg$ excess at 750 GeV and the $h\to\mt$ excess are due to new physics.
A minimal way to introduce  (renormalisable) flavour violation
and extra bosons to the Standard Model  is to add a second Higgs doublet.
Its $\tau \leftrightarrow \mu$ coupling may be connected to large $2-3$ mixing in the neutrino sector,
in scenarios where the Yukawa couplings of charged leptons and neutrinos are related.

The neutral scalars $H$ and $A$ can play the role of the $750$ GeV resonance, even though the strength of the excess in the early 13 TeV data
is significantly larger than the one expected in the 2HDM alone. We take this as a hint that additional states close to the TeV are
present in the underlying theory, with large Yukawa couplings to the second Higgs doublet. We have shown that a pair of vector-like fermions is sufficient
to reproduce the right cross-section, and respect all other constraints.
However, such fermions must have gauge charges larger than the Standard Model fermions: indicatively, for a Yukawa $\simeq 3$ and $R_c\le 8$, one needs
$|Q|\ge 2$ in scenarios (A) and (B), and $|Q|\ge 5$ in scenario (C), see section 4.
Alternatively, several pairs of fermions have to be introduced.
These are important indications to constrain those well-motivated extensions of  the Standard Model 
that predict vector-like fermions, such as top partners.

Were the  heavy Higgses to have no couplings
to Standard Model fermions, then  
$gg \to H,A \to \gg$ is a natural discovery
channel. However, to explain $\htm$,
the heavy Higgses must interact with $\tau^\pm
\mu^\mp$, 
and mixing is required between $h$ and $H$.
Both requirements  gives Standard Model decay channels to
$H$ and $A$, which  reduces  $BR(H,A \to gg,\gg)$;
nonetheless we find three scenarios that fit both
excesses. In addition,  the mixing  must respect both a lower bound 
to reproduce the LFV excess, and an upper 
bound to protect the 125 GeV Higgs couplings:
$10^{-3}\lesssim\cbma\lesssim 0.3$.

The decay $\tmg$ is a particular challenge for
this model, because  the
heavy Higgses couple to  $\tau^\pm
\mu^\mp$ and have  an enhanced coupling to $\gg$.
In combination, these interactions give
a  ``Barr-Zee''  contribution to   $\tmg$
which is dangerously large.  By
choosing the Yukawas to obtain destructive interference
between  $A$ and $H$, we find that at least two of the
 scenarios are compatible  with
the current experimental limit on $\tmg$.

\section*{Acknowledgements}

We warmly thank Felix Br\"ummer for several enlightening discussions.
We acknowledge the partial support of the OCEVU
Labex (ANR-11-LABX-0060) and the A*MIDEX project (ANR-11-IDEX-0001-02),
funded by the ``Investissements d'Avenir" French government program managed by the ANR.
MF acknowledges the partial support of the  European Union FP7  ITN INVISIBLES (Marie Curie Actions, PITN-GA-2011-289442).

\bibliographystyle{JHEP}
\providecommand{\href}[2]{#2}\begingroup\raggedright\endgroup

\end{document}